\documentclass[runningheads]{llncs}
\usepackage[T1]{fontenc}
\usepackage{amsmath}
\usepackage{amssymb}
\usepackage{url}
\usepackage{xurl}
\usepackage{graphicx}
\usepackage{lineno}
\usepackage{microtype}
\usepackage{subcaption}
\usepackage{booktabs}
\usepackage{xcolor}
\usepackage{hyperref}

\newcommand{\fixme}[1]{\textcolor{black}{#1}}

\begin{document}
%

\title{Robot Visions: Breaking reCAPTCHA at \\Zero Cost and Zero Shot}

\titlerunning{Robot Visions: Breaking reCAPTCHA}
%
\author{Suphannee Sivakorn \and
Samantha Gottlieb
}


%

\authorrunning{S. Sivakorn and S. Gottlieb}
%

\institute{Independent Researchers \\
\email{\{ssivakorn,sagottlieb\}@gmail.com}
}

\maketitle              
%

\begin{abstract}

Google reCAPTCHA is the most widely deployed visual CAPTCHA service,
protecting hundreds of thousands of websites from automated bots. It
serves as a critical line of defense against automated attacks, including
credential stuffing, bulk account creation, and automated form abuse.
It has proven largely effective since its introduction in 2007. However,
the rise of accessible AI now threatens its efficacy.

Prior work has demonstrated that commercial cloud-based vision-language models
(VLMs) can solve visual CAPTCHA challenges, but at non-trivial monetary
cost per attempt.
In this paper, we show that \emph{free} and \emph{locally-run} models
can break Google reCAPTCHA.
We conduct a comprehensive study of reCAPTCHA and present a taxonomy of
its challenge types: Type~A (independent image tiles, with static and
dynamic sub-variants) and Type~B (a single image partitioned into a
$4{\times}4$ grid), each demanding a distinct solving strategy.

We design zero-shot, no-cost solvers built entirely on \emph{open-source}
local models, specifically CLIP (58\% \fixme{per-challenge} accuracy on Type~A) and OWLv2
(43.5\% on Type~B), requiring no model training and no API access. Our
end-to-end automated solver achieves a \textbf{92.6\%} \fixme{per-session} success
rate across 500 real-world reCAPTCHA sessions.
We further demonstrate that reCAPTCHA can be defeated
by a non-technical adversary, using only natural-language instructions to a
commodity AI assistant. This collapses the practical attacker skill floor to
near zero and fundamentally changes the threat model for challenge-based
CAPTCHAs.
Although reCAPTCHA
increasingly favors reputation-based verification, visual challenge-based
fallback persists as a safety net that, paradoxically, has become the
weakest link in the defense chain, suggesting that challenge-based visual
CAPTCHAs may have reached the end of their useful life.



\keywords{CAPTCHA \and reCAPTCHA \and visual CAPTCHA \and vision models \and bot detection \and AI agent \and AI \and security}
\end{abstract}

%
\section{Introduction}
\label{sec:intro}

CAPTCHAs (Completely Automated Public Turing tests to tell Computers and
Humans Apart) are a cornerstone of web security, designed to distinguish
human users from automated bots. Their importance has grown considerably
in the age of AI, where autonomous systems increasingly perform
actions that were once assumed to require human presence. Bots now
account for over half of all global internet traffic, and autonomous AI
agents are rapidly taking over web browsing, data scraping, and online
transactions~\cite{thales2026badbots}.
This has fueled large-scale credential stuffing, content
scraping, and malicious account takeovers. The threat is compounded by
agentic AI systems capable of discovering vulnerabilities and chaining
them into full attack pipelines~\cite{glasswing2025,ft2025ai}. A
concrete example is the bulk registration of malicious domain names,
which serve as infrastructure for domain generation algorithms (DGAs)~\cite{icann2023},
phishing campaigns, drive-by downloads, and command-and-control (C2)
networks~\cite{falconfeeds2024}, many of which depend on automated
web-based interactions that CAPTCHAs are designed to block.
At the same time, CAPTCHAs impose a usability cost. It is widely
accepted that users consider CAPTCHAs a nuisance, and even simple
challenges deter a significant fraction of legitimate users from
completing their intended actions~\cite{cloudflare2022}. The challenge
is therefore twofold: solving a CAPTCHA must remain effortless for
humans while staying robust against automated solvers.

Among all CAPTCHA services, Google reCAPTCHA is by far the most
widely deployed: over
a quarter of the top one million websites use CAPTCHAs, of which 94\% employ visual challenges~\cite{halligan2025}.
%
%
The security assumption underlying visual CAPTCHAs is that image
understanding tasks are \emph{bot-hard} and \emph{human-easy}. This
assumption has been increasingly challenged. Sivakorn et
al.~\cite{sivakorn2016} demonstrated in 2016 that deep learning
models could solve Google reCAPTCHA at 70.78\% accuracy using
CNN-based semantic image annotation. More recently, Teoh et
al.~\cite{halligan2025} showed that a generalized VLM agent
(Halligan) using the GPT-4o cloud API achieves 60.7\% accuracy
across 26 CAPTCHA types at \$0.024 per challenge.

However, several important questions remain open. Can free,
open-source, locally-run vision models break Google reCAPTCHA without
any API cost or training? And can it be broken
with \emph{zero lines of code}, simply by providing natural-language
instructions to an AI assistant?

In this paper, we answer both questions affirmatively. Our
contributions are:

\begin{enumerate}



    \item \textbf{\fixme{Zero-shot, zero-cost solvers.}} We solve each challenge type
    with an open-source, locally-run model that requires no training, no labeled
    data, and no API access: CLIP~\cite{radford2021clip} for Type~A (best accuracy 58\%) and
    OWLv2~\cite{minderer2023owlv2} for Type~B (43.5\%). \fixme{Unlike prior attacks that
    fine-tune fixed-class detectors on labeled CAPTCHA data~\cite{10633630} or query paid
    cloud vision-language models~\cite{halligan2025}, our solvers are \emph{zero-shot} and
    \emph{open-vocabulary}: they generalize to an arbitrary target keyword with no
    fine-tuning and no predefined class list.} We define a model
    as \emph{locally-run} in the sense relevant to this work if it executes
    on commodity hardware without a dedicated GPU and without per-query
    monetary cost. This excludes cloud-served models (e.g., GPT-4o,
    Gemini) regardless of license, as well as larger open-source
    multimodal models (e.g., LLaVA, InternVL) whose practical variants
    require substantial GPU memory unavailable on typical consumer devices.
    An end-to-end pipeline built on these models achieves a 92.6\% session
    success rate across 500 real-world reCAPTCHA sessions.

    \item \textbf{Zero-Barrier Attack Surface.} We show that reCAPTCHA can be defeated by a non-technical adversary, using only natural-language instructions to a commodity AI assistant. This collapses the attacker skill floor to near zero and fundamentally changes the threat model for challenge-based CAPTCHAs.
\end{enumerate}



\section{Background}
\label{sec:background}

\subsection{Google reCAPTCHA v2 and v3}
Google reCAPTCHA is available in two major versions, both built on a
shared risk analysis system that evaluates browser characteristics,
cookies, and browsing history to assign a reputation score to each
session~\cite{sivakorn2016,friendlycaptcha2026}. Verified users with
high reputation scores may bypass challenges entirely, while unknown or
suspicious sessions are systematically escalated to a visual image
challenge.


\textbf{reCAPTCHA v2}~\cite{recaptchav2} presents a checkbox labeled
``I'm not a robot.'' Users with high trust scores receive the checkmark
immediately upon clicking. For lower-confidence requests, a visual image
challenge is presented, requiring users to select all tiles matching a
keyword (e.g., ``fire hydrant''). Multiple challenges may
be presented in sequence before the CAPTCHA is fully solved. A session
expires after approximately \emph{two minutes} without a submission, at which
point the checkbox resets and the user must restart from scratch.
We identify two distinct challenge types, which we term Type~A and Type~B, described
in Section~\ref{sec:challenge-types}.


\textbf{reCAPTCHA v3}~\cite{recaptchav3} is fully invisible. It
analyzes behavioral signals in the background and returns a continuous
risk score between 0.0 (highly suspicious) and 1.0 (highly likely
human), without ever presenting a challenge to the user. Site
administrators configure their own acceptance thresholds based on this
score. When v3 cannot confirm a user with sufficient confidence, a
visual challenge is deployed as a fallback before
activities requiring user verification, as defined by the site
administrator~\cite{friendlycaptcha2026}. Ironically, it is precisely
this fallback mechanism, designed to catch what behavioral analysis
misses, that we demonstrate is solvable by zero-cost, open-source
visual AI models.

\begin{figure}[tb]
\centering
\includegraphics[width=0.4\columnwidth]{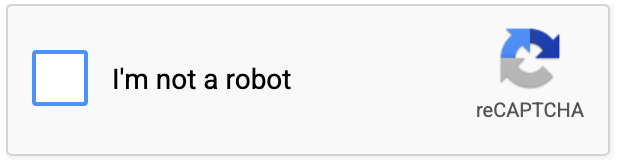}\quad
\includegraphics[width=0.4\columnwidth]{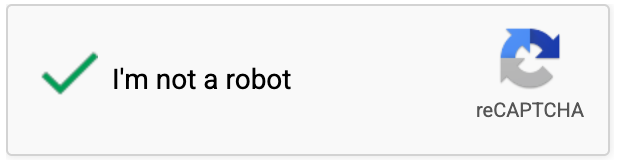}
\caption{reCAPTCHA v2 checkpoint. \textbf{Left:} the initial checkbox prompt. \textbf{Right:} checkmark state after successful challenge completion.}
\label{fig:recaptcha-intro}
\end{figure}

\textbf{Role of User Reputation.}
A key but underappreciated aspect of reCAPTCHA v2 and v3 is that
challenge difficulty, and whether a challenge is shown at all, depends
heavily on the user's Google account reputation and browsing history.
Sivakorn et al.~\cite{sivakorn2016} demonstrated that signed-in Google
users with high reputation scores can bypass the visual challenge
entirely, receiving an immediate checkmark without answering any
images, with the reputation signal encompassing account age, browsing
history, and prior CAPTCHA performance.

This reputation effect can skew experimental results, since a bot may
pass reCAPTCHA simply because its browser is signed in with a trusted
Google account or carries existing browsing history, rather than
because it solved the image challenge correctly. To avoid this bias,
all experiments in this paper use fresh, not-signed-in browser sessions,
ensuring reCAPTCHA consistently presents visual challenges. Controls
are detailed in Section~\ref{sec:exp-isolation}.



\subsection{reCAPTCHA Visual Challenge Types}
\label{sec:challenge-types}

\begin{figure}[tb]
\centering
\includegraphics[width=0.3\columnwidth]{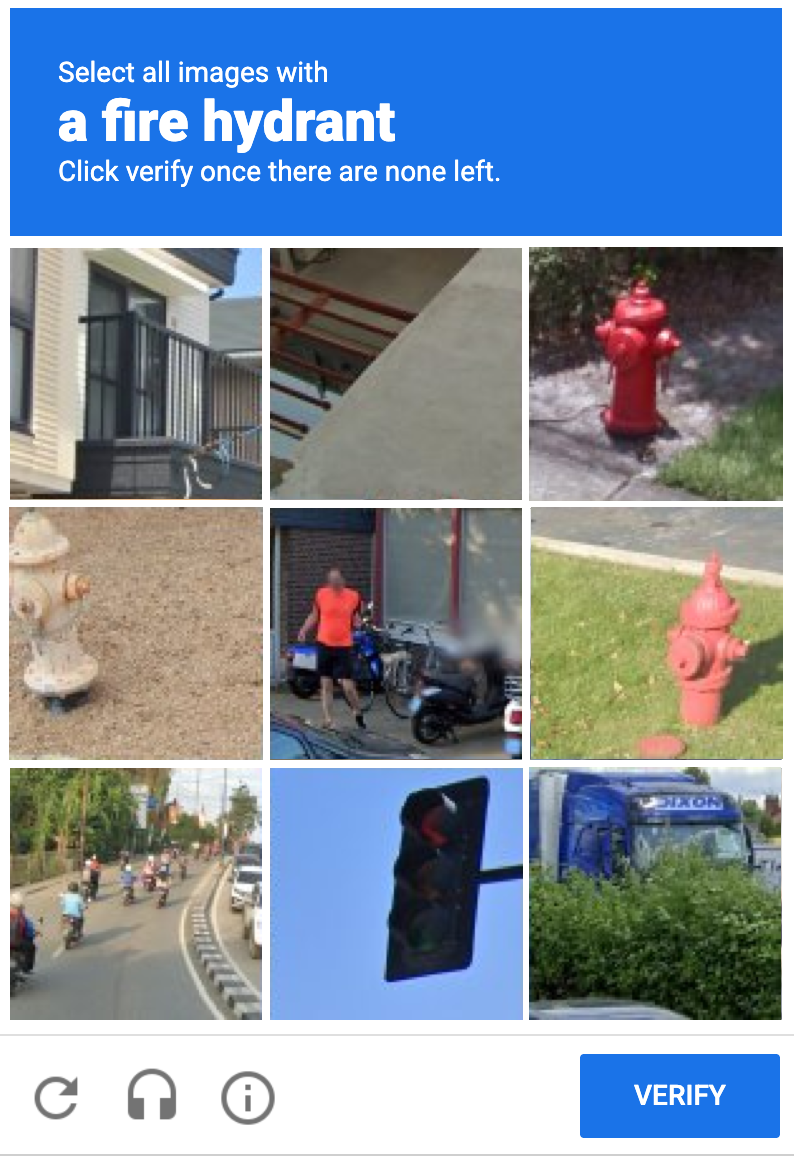}\quad
\includegraphics[width=0.3\columnwidth]{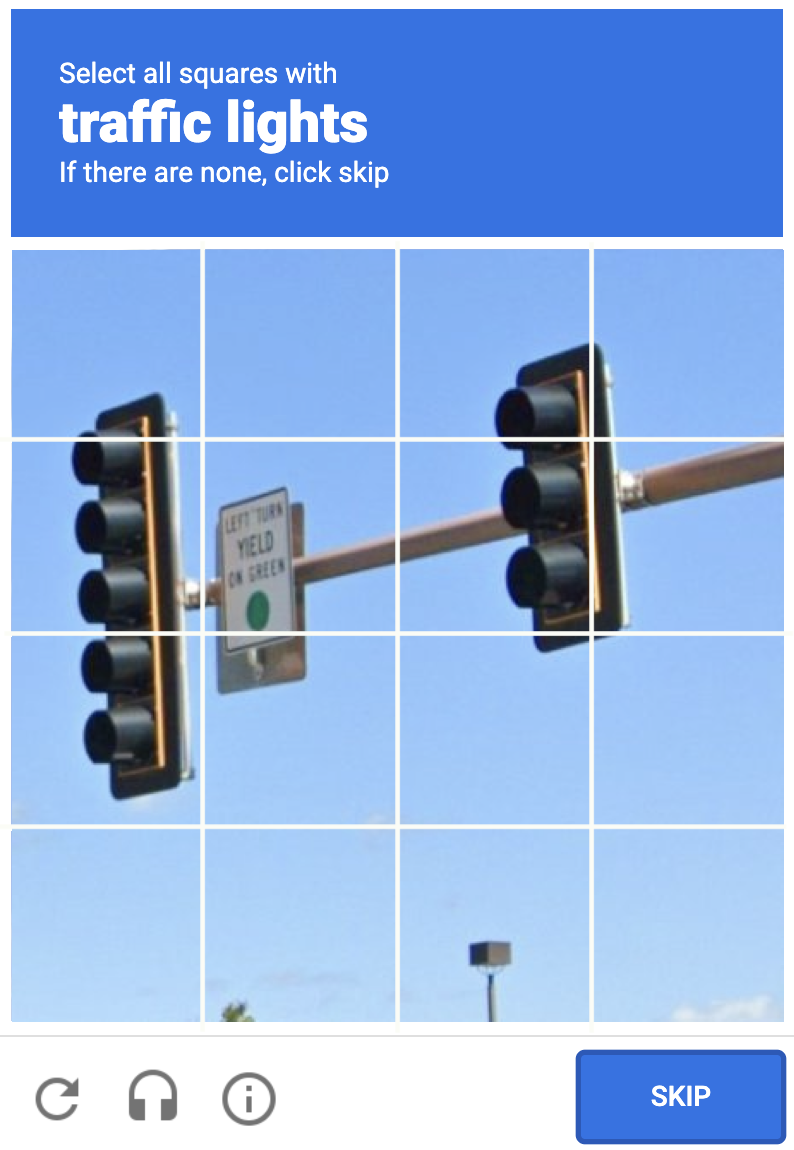}
\caption{reCAPTCHA visual challenge examples. \textbf{Left:} Type~A (keyword: ``a fire hydrant''). \textbf{Right:} Type~B (keyword: ``traffic lights'').}
\label{fig:challenge_types}
\end{figure}

Google reCAPTCHA presents two distinct challenge types that require
fundamentally different solving strategies\fixme{, a distinction also
drawn in prior work~\cite{10633630}}, as illustrated in
Figure~\ref{fig:challenge_types}.

\begin{figure}[tb]
\centering
\includegraphics[width=0.28\columnwidth]{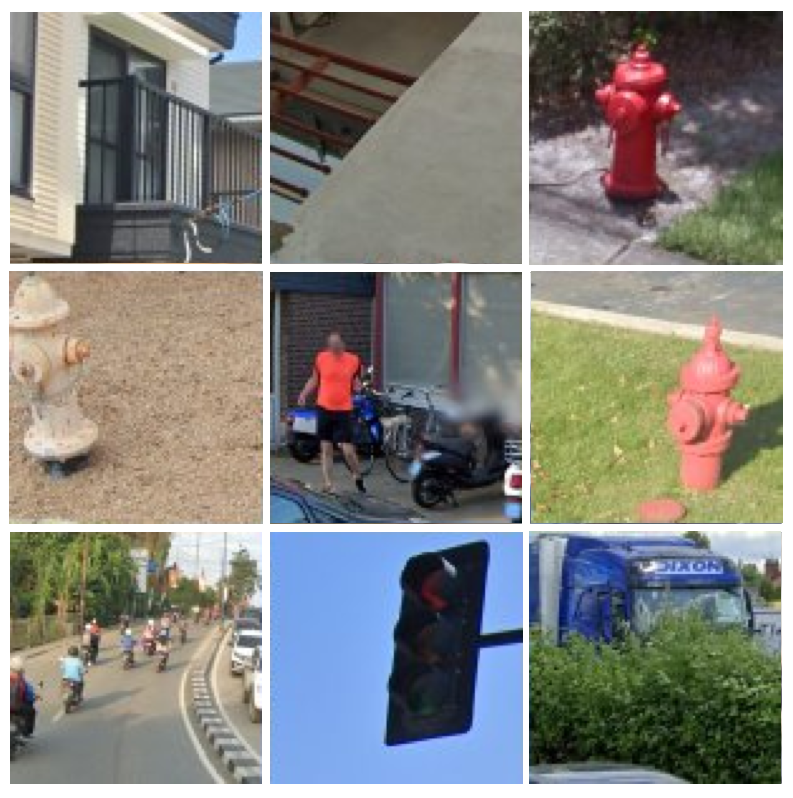}\hfill
\includegraphics[width=0.28\columnwidth]{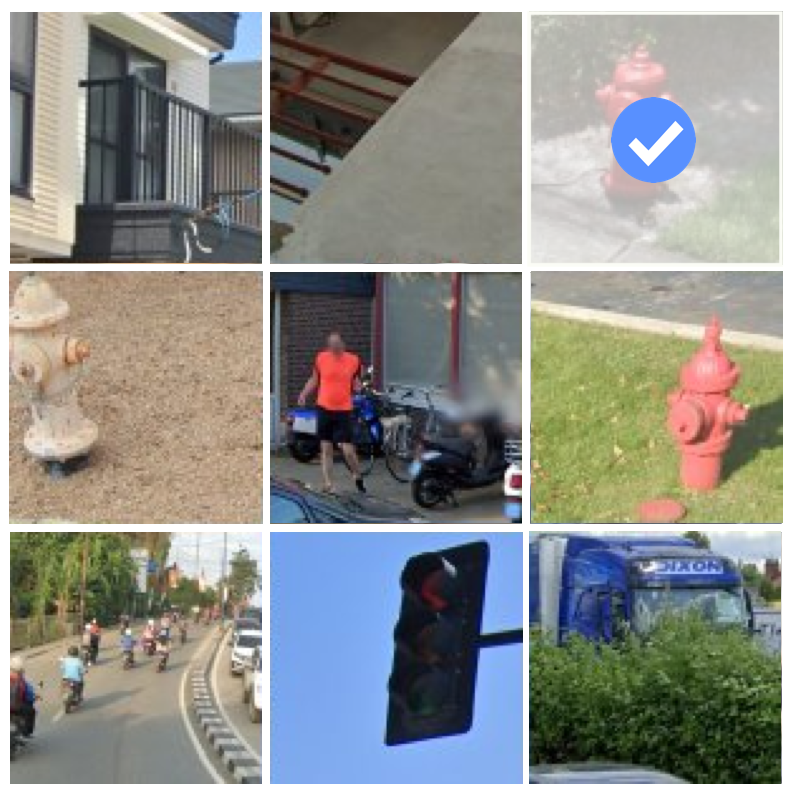}\hfill
\includegraphics[width=0.28\columnwidth]{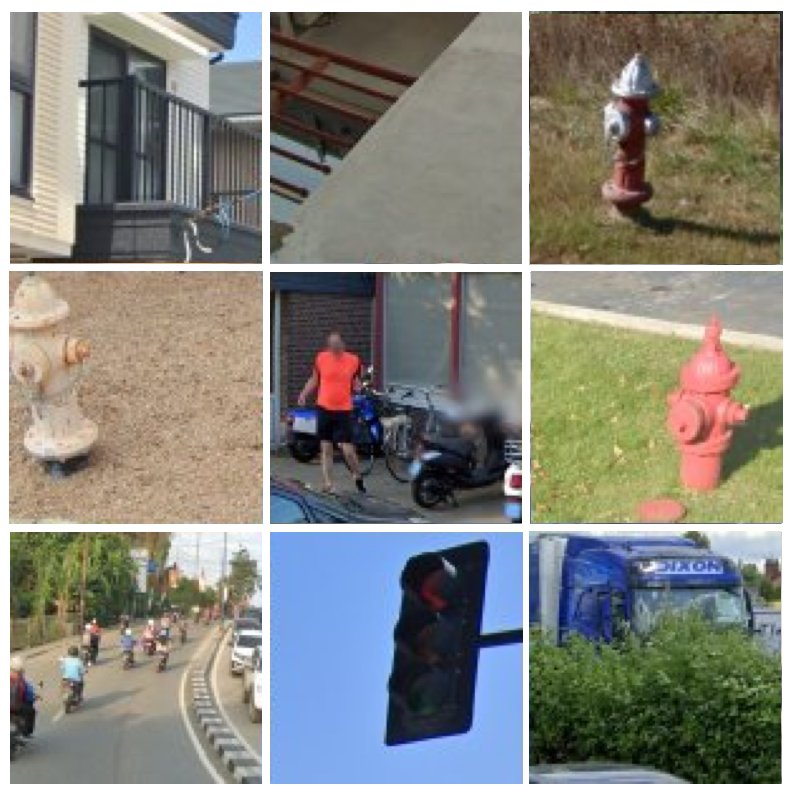}
\caption{Type~A (keyword: ``a fire hydrant''). \textbf{Left:} nine independent images in a 3×3 grid. \textbf{Center and Right:} clicking a correct tile (containing a fire hydrant) causes it to refresh with a new random image.}
\label{fig:hydrant-typeA}
\end{figure}

\textbf{Type~A: Independent Images per Tile.} The user receives a 3×3 grid of nine tiles and is instructed to ``Select all images with [keyword].'' All nine tiles share a single sprite image, with CSS offsets making each tile show a different crop. Type~A comes in two sub-variants: (1) \emph{Dynamic} challenges include the subtitle ``Click verify once there are none left''. Clicking a correct tile replaces it with a new random image, and the challenge loops until no matches remain. (2) \emph{Static} challenges have no subtitle. Clicking a tile only adds a selection border without refreshing the image, so all matches must be identified in a single pass. Figure~\ref{fig:hydrant-typeA} shows a dynamic example with keyword ``a fire hydrant.''

\begin{figure}[tb]
\centering
\includegraphics[width=0.28\columnwidth]{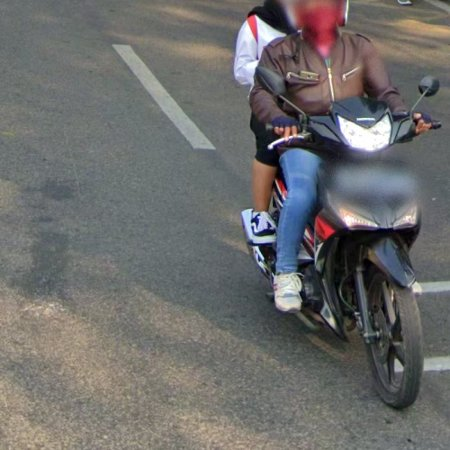}\hfill
\includegraphics[width=0.28\columnwidth]{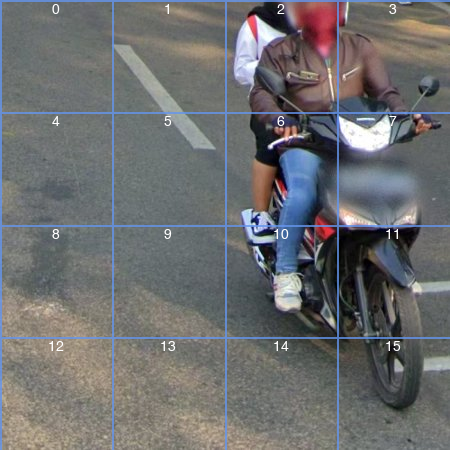}\hfill
\includegraphics[width=0.28\columnwidth]{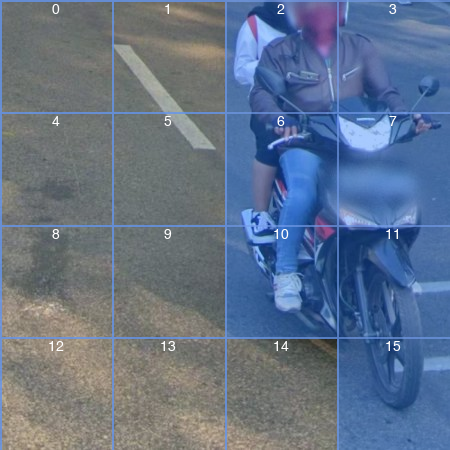}
\caption{Type~B (keyword: ``motorcycles''). \textbf{Left:} original
image showing a motorcycle on a road. \textbf{Center:} image partitioned into a
$4{\times}4$ tile grid. \textbf{Right:} tile grid with correct answer tiles
highlighted.}
\label{fig:motorcycles-typeB}
\end{figure}

\textbf{Type~B: Single Image Split into Tiles.} The user receives a 4×4 grid of 16 tiles, each a crop of the same underlying photograph, and is instructed to ``Select all squares with [keyword].'' Unlike Type~A, there is only a static version and all matching tiles must be identified in a single pass. Figure~\ref{fig:motorcycles-typeB} shows an example challenge with keyword ``motorcycles.'' The original image (left) depicts motorcycles on a road; the tiles are spatial crops that may include partial objects, surrounding context, and background clutter (e.g., road and shadows).


\subsection{Image CAPTCHA Solving}
Early attacks on image-based CAPTCHAs relied on classical computer vision
techniques such as edge detection, segmentation, and OCR to break
distorted-text challenges~\cite{bursztein2011text}. As CAPTCHA designs
evolved toward semantic image recognition tasks, attack strategies shifted toward deep learning.
Sivakorn et al.~\cite{sivakorn2016} presented the first deep learning
attack on Google reCAPTCHA, achieving 70.78\% accuracy using CNN-based
semantic image annotation at 19 seconds per challenge. They also analyzed
reCAPTCHA's risk analysis system and browser fingerprinting mechanisms,
revealing how reputation signals influence challenge presentation.
Subsequent work applied increasingly powerful vision models to CAPTCHA
solving. Specialized solvers have leveraged image classifiers~\cite{10.3389/fnins.2017.00309,ILSVRC15},
object detectors such as YOLO~\cite{redmon2016yolo}, and visual question
answering models, each requiring task-specific training or fine-tuning on
CAPTCHA datasets~\cite{8887344,8665729}.
\fixme{%
Most directly related to our work, Plesner et al.~\cite{10633630} broke
reCAPTCHA v2 using YOLOv8 and reported near-perfect solve rates. For the
tile-selection challenges, which correspond to our Type~A, they fine-tuned a YOLOv8
classification model on approximately 14k labeled images spanning a fixed set of 13 object classes. For the single-image grid challenges,
which correspond to our Type~B, they applied a pretrained YOLOv8 segmentation model
without further fine-tuning. Their Type~A tile classifier is therefore bound to
the 13 object classes it was trained on and cannot generalize to target
objects outside that fixed set, whereas our zero-shot CLIP classifier accepts
an arbitrary keyword with no training and no predefined class list.}
Most recently, Teoh et al.~\cite{halligan2025} introduced Halligan, a
generalized VLM agent that solves 26 CAPTCHA types using the GPT-4o cloud
API at \$0.024 per challenge and 60.7\% accuracy. While broader in scope,
Halligan relies on a commercial cloud API. Our work builds directly on this foundation,
extending it to zero-shot local models that require no training and operate at zero cost.

Across this body of prior work, every solver depends on at least one of the
following: task-specific training with labeled data, \fixme{a fixed set of predefined
object classes}, or paid cloud-service APIs. We are the first to demonstrate
that free, open-source, locally-run zero-shot and open-vocabulary models,
requiring no training, no labeled data, no predefined class list, and no API
access, can break Google reCAPTCHA at scale.

\subsection{AI Vision-Language Models}
Vision-language models (VLMs) learn joint representations of images and
text, enabling them to perform image understanding tasks guided by natural
language without task-specific training. This zero-shot capability makes
them particularly relevant to CAPTCHA solving.

\textbf{CLIP (Contrastive Language--Image Pre-training)}~\cite{radford2021clip}
established a foundation for zero-shot image classification by training
on 400 million image-text pairs using a contrastive objective. Given an
image and a set of text labels, CLIP scores their similarity without any
fine-tuning, enabling flexible classification across arbitrary categories.
This makes CLIP well-suited to Type~A challenges, where each tile is an
independent image requiring per-tile classification. CLIP follows the
Vision Transformer (ViT) architecture~\cite{dosovitskiy2020vit}, with
model names encoding backbone size (B: Base, L: Large, H: Huge) and
patch size in pixels (e.g., ViT-H/14 uses the largest backbone with
$14\times14$ patches). Larger backbones yield richer representations at
higher inference cost; smaller patches capture finer detail at higher
computational cost.

\textbf{OWLv2 (Open-World Localization v2)}~\cite{minderer2023owlv2}
extends zero-shot capabilities to object detection by localizing and
classifying objects within an image based on arbitrary open-vocabulary
text queries. Unlike CLIP, which operates solely at the global image
level, OWLv2 yields bounding boxes and confidence scores for individual
object instances, making it well-suited to Type~B challenges, where the
solver must determine which tiles of a single image a target object
occupies. The model projects visual and linguistic features into a
shared embedding space, computing similarity scores between image
patches and text prompts; a localized region with high similarity to a
text query (e.g., ``dog'') is detected even if the model was never
explicitly trained on that category.

\textbf{Other Vision-Language Models (VLMs).}
Several powerful VLMs exist that are well-suited to visual reasoning
tasks but do not meet the zero-cost, locally-run constraints of this
work. GPT-4o~\cite{openai2024gpt4o}, GPT-5~\cite{openai2025gpt5}, and
Gemini~\cite{google2024gemini} are state-of-the-art multimodal models
capable of fine-grained image understanding, but require paid cloud
API access per query~\cite{halligan2025}.

LLaVA~\cite{liu2023llava} and InternVL~\cite{chen2024internvl} are
open-source multimodal models that can run locally, but their smallest
practical variants start at 7B and 8B parameters, respectively,
roughly 10$\times$ larger than CLIP ViT-H/14 (632M) and OWLv2
(900M), requiring substantial GPU memory that makes them
inaccessible on commodity hardware without dedicated GPUs.
Furthermore, VLMs are not
optimized for image classification (Type~A) and object localization (Type~B)
tasks that reCAPTCHA demands, offering no advantage over our
lightweight zero-shot models for this specific task.

Our work deliberately constrains itself to lightweight, zero-cost models
to demonstrate that the barrier to breaking reCAPTCHA does not require
frontier model access. Together, the CLIP and OWLv2 models cover the distinct visual reasoning demands of
each reCAPTCHA challenge type without requiring any fine-tuning, heavy compute resources, or commercial API access.

\subsection{Agentic AI for Offensive Security}
Large language models are demonstrating increasing
capability to assist in and autonomously execute cyber attacks.
Models such as GPT-4o~\cite{openai2024gpt4o},
GPT-5~\cite{openai2025gpt5}, and Claude models~\cite{claude_sonnet}
can generate functional code from natural-language descriptions alone~\cite{chen2021codex}. In the security
domain, this capability fundamentally lowers the barrier to entry for
constructing attack tooling: an adversary no longer needs programming
expertise to build a working exploit. PentestGPT~\cite{deng2024pentestgpt}
demonstrated that LLM-guided agents can perform penetration testing
tasks by reasoning about system state and chaining reconnaissance,
exploitation, and post-exploitation steps. More recently, frontier AI
systems have been shown to autonomously exploit vulnerabilities
in real-world systems~\cite{fang2024llmagents}, and Anthropic's
Project Glasswing~\cite{glasswing2025} highlights growing concern
around agentic AI systems chaining vulnerabilities into full
exploitation pipelines. Our work contributes to this landscape: our
end-to-end reCAPTCHA solver pipeline was generated via AI-assisted
coding, and our prompt-only solver requires no code whatsoever,
demonstrating that CAPTCHA-breaking attack tooling is now accessible
to anyone with an AI assistant.

\section{Methodology}
\label{sec:method}
In this section, we present an overview of our system
designed to solve reCAPTCHA challenges. Our system is built on Selenium~\cite{selenium}, an open-source browser
automation framework. We opt for the ChromeDriver~\cite{chromedriver},
which lets us leverage the functionality of the browser
engine and handle all aspects of the web pages required for
bypassing the browser checks of reCAPTCHA. The WebDriver offers
functionality for locating specific HTML DOM elements and provides features for handling mouse events.

\subsection{Experimental Isolation}
\label{sec:exp-isolation}

To isolate the image-classification capability of our solver from user-reputation effects, all experiments in this paper are conducted under controlled conditions:

\begin{itemize}
  \item \textbf{No Google sign-in.} All sessions are run with a fresh browser profile not signed in to any Google account, ensuring no account reputation contributes to the outcome.
  \item \textbf{No prior browsing history.} Each session starts from a clean browser state. No other websites are visited before or during the reCAPTCHA session, avoiding residual trust signals from browsing patterns.
  \item \textbf{New browser per session.} Each experimental session launches a new browser instance to prevent cookie or fingerprint persistence across sessions.
\end{itemize}

Under these conditions, reCAPTCHA consistently presents visual image
challenges rather than granting immediate checkmarks. This represents
the hardest-case scenario for automated solvers, the same conditions
faced by bots with no user history, and ensures our accuracy
measurements reflect genuine image-classification performance rather
than reputation-based bypasses.

\subsection{Challenge Data Collection}
\label{sec:data-collection}

To study these challenge types and develop our challenge solver, we used
an automated Selenium-based crawler to collect 1,000 labeled challenges
(662 Type~A and 338 Type~B) from various websites that have deployed
Google reCAPTCHA.

\textbf{Navigation and Challenge Extraction.}
Google reCAPTCHA is embedded within two nested iframes that Selenium
must navigate explicitly. The crawler first locates and switches into
the checkbox iframe (\texttt{iframe[title="reCAPTCHA"]}), then clicks
the checkbox border element to trigger the visual challenge. It then
switches into the challenge iframe (\texttt{iframe[src*="bframe"]})
where all further DOM interaction takes place.
Inside the challenge iframe, the keyword is extracted from the
\texttt{<strong>} element within \texttt{div.rc-imageselect-instructions},
and tiles are located as \texttt{div.rc-image-tile-target} elements. For
each challenge, the crawler records (1)~the full instruction text shown
to the user (e.g., ``Select all images with bicycles''), (2)~the
extracted keyword (e.g., ``bicycles''), and (3)~the full challenge
image (if available) and individual tile images, one per tile.

\textbf{Manual Annotation.}
Each collected challenge was manually answered by a human annotator
using a custom web interface, recording which tiles constitute correct
answers. This offline labeled dataset is not used for training, but rather to
select the best visual models and tune parameters for each challenge
type (Section~\ref{sec:eval_offline}).

The dataset spans 13 distinct keywords observed in the wild.
Transportation-related objects dominate, with bicycles (18.3\%), buses
(17.2\%), crosswalks (15.7\%), motorcycles (13.3\%), and cars (11.9\%)
collectively accounting for over 76\% of all challenges. The remaining
keywords (i.e., fire hydrants, traffic lights, stairs, taxis, tractors, and
bridges) appear less frequently.

\subsection{Offline Challenge Solver}

We design two challenge solvers
that address the fundamentally different visual reasoning demands of
Type~A and Type~B challenges: (1) Local Vision Models and (2) Prompt-only Skill.

\subsubsection{Local Vision Models.}~We employ two free, open-source, locally-run vision models, each
selected for its suitability for a specific challenge type. Both models
can run on a GPU but do not require one, making them accessible on
commodity hardware without external services. All experiments in this paper
were conducted on a consumer laptop (integrated GPU, no dedicated GPU),
demonstrating that the solver requires no specialized hardware.


\textbf{CLIP for Type~A.} We apply CLIP to Type~A challenges via zero-shot
per-tile classification. For each tile, CLIP computes \emph{cosine similarity} against two text prompts:
a \emph{positive prompt} ``a photo of \{keyword\}''
and a negative prompt ``a photo of something else.'' The classification score is defined as the difference between
the two similarities:
$ \text{score} = \text{sim}_{\text{pos}} - \text{sim}_{\text{neg}}$.

A tile is selected if its score meets or exceeds a tuned \emph{threshold}.
This contrastive formulation is more discriminative than using raw
positive similarity alone: a tile that weakly resembles the keyword
but more strongly resembles ``something else'' receives a negative score
and is correctly rejected. Scores typically range from $-0.10$ to
$+0.10$, so small threshold differences of 0.01 to 0.05 meaningfully
affect recall and precision.
Table~\ref{tab:clip-example} illustrates the scoring for keyword
``cars'' at threshold 0.02. Note that Tile~2 scores positively yet is
correctly rejected, as its score falls below the threshold.
We evaluate three model sizes: ViT-B/32, ViT-L/14, and ViT-H/14,
with per-model thresholds tuned on the offline dataset
(Section~\ref{sec:eval_offline}).

\begin{table}[tb]
\centering
\caption{Example CLIP scoring for keyword ``cars'' (\emph{threshold} $ = 0.02$).}
\label{tab:clip-example}

\scriptsize
\begingroup
\setlength{\tabcolsep}{4pt}
\begin{tabular}{llrrrl}
\toprule
Tile \#& Image Content & $\text{sim}_{\text{pos}}$ & $\text{sim}_{\text{neg}}$ & Score & Decision \\
\midrule
Tile 0 & car      & 0.34 & 0.28 & +0.06 & \checkmark~selected \\
Tile 1 & building & 0.27 & 0.29 & $-$0.02 & $\times$~rejected \\
Tile 2 & road     & 0.28 & 0.27 & +0.01  & $\times$~rejected \\
... &  &  &  &  & \\
Tile 8 & car+road & 0.32 & 0.28 & +0.04 & \checkmark~selected \\
\bottomrule
\end{tabular}
\endgroup
\end{table}

\textbf{OWLv2 for Type~B.} Type~B challenges require reasoning about
spatial layout: the solver must determine not only whether an object is
present, but which tiles it occupies. Per-tile classification is
insufficient here, as individual tile crops may contain only a partial
object or ambiguous background. We confirm this empirically: CLIP
applied directly to Type~B achieves only 12.7\% accuracy, compared to
43.5\% for OWLv2, a 3.4$\times$ improvement. We therefore apply OWLv2
on the full image and map detected bounding boxes to tiles via
intersection-over-tile-area coverage.

\begin{figure}[tb]
\centering
\includegraphics[width=0.28\columnwidth]{figures/motorcycles_original.png}\hfill
\includegraphics[width=0.28\columnwidth]{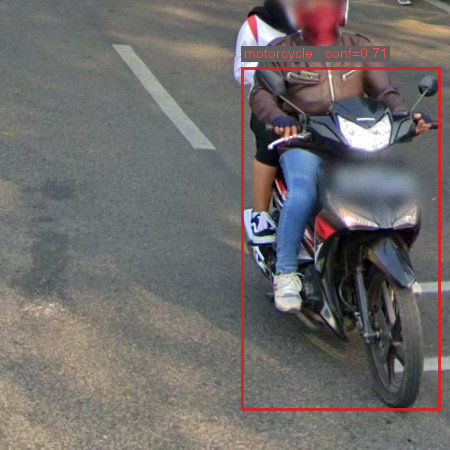}\hfill
\includegraphics[width=0.28\columnwidth]{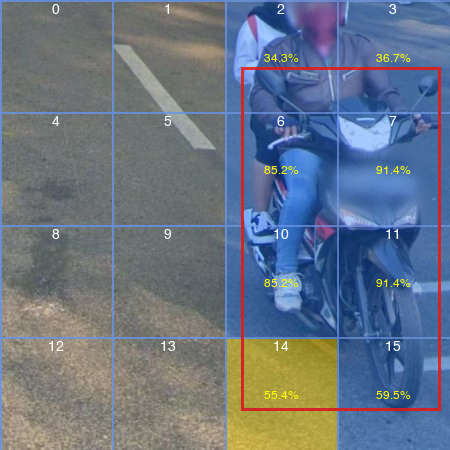}
\caption{\textbf{Left:} Original challenge image (keyword: ``motorcycles'').%
~\textbf{Center (Stage~1):} OWLv2 detects the target
object on the full image, producing an axis-aligned bounding box
(confidence $= 0.71$).%
~\textbf{Right (Stage~2):} The bounding box is
mapped onto the $4{\times}4$ tile grid; tiles whose overlap exceeds the
threshold ($\geq 10\%$) are selected (blue). Tile~14 (yellow, 55.4\%
overlap) is a false positive: the bounding box extends below the
motorcycle body into the road surface, inflating coverage for a tile
with no target object.}

\label{fig:owlv2-stages}
\end{figure}

Tile selection proceeds in two stages. First, OWLv2 detects objects on
the full image, producing bounding boxes with confidence scores between
0 and 1; boxes below the confidence threshold are discarded. Second,
each surviving box is mapped onto the tile grid, and tiles covered by
more than the coverage threshold are selected. These two thresholds
guard against distinct failure modes: confidence prevents false
detections, while coverage prevents selecting tiles with only marginal
overlap. Figure~\ref{fig:owlv2-stages} illustrates this process: Stage~1
shows a detection with confidence 0.71, and Stage~2 shows the resulting
tile coverage. Tile~14 receives 55.4\% coverage, exceeding the 10\%
threshold and is selected, but this is a false positive: the bounding
box extends below the motorcycle body into the road surface, inflating
coverage for a tile with no target object.

We also experimented with a hybrid approach incorporating the Segment
Anything Model (SAM)~\cite{kirillov2023sam} to refine OWLv2's bounding
boxes into precise segmentation masks, hypothesizing that pixel-level
masks would reduce the coverage ambiguity caused by coarse bounding
boxes. In practice, SAM yielded only a 1--2\% accuracy improvement
over OWLv2 alone, while increasing inference time substantially. Given
this unfavorable cost-benefit trade-off, we exclude SAM from our final
pipeline and report OWLv2-only results throughout.


\subsubsection{Prompt-Only Skill.}
\label{sec:prompt-only-offline}
~The Local Vision Models approach requires downloading model weights and running local inference.
To assess how far a completely zero-setup approach can reach, we
implement a \emph{prompt-only} solver that requires no model,
no coding, and no external API key. This solver also
serves as a direct measure of the visual reasoning capability of a
frontier multimodal language model on the reCAPTCHA task.

The solver is realized as a \emph{skill}, a structured markdown
file of natural-language instructions consumed by the Claude Code
AI assistant~\cite{claude_code} and
invoked by a command, \texttt{/challenge-solver}. The skill
instructs the assistant in three steps. First, it reads the
instruction text to obtain the keyword and detect the challenge
type (``images'' $\to$ Type~A; ``squares''
$\to$ Type~B; fallback: tile count). Second, it reads the tile images
using its built-in file-reading tool and applies its \emph{multimodal vision}~\cite{claude_multimodal}
capability directly. For Type~A, each of the nine
\texttt{tile\_N.png} files is read; \fixme{the agent first lists the
concrete objects visible in the tile, independent of the keyword, then judges whether the
keyword object is present and selects the tiles.} \fixme{For
Type~B, the agent estimates the percentage of each tile region covered by the
keyword object and selects any tile with nonzero coverage (a threshold of 0\%, i.e.,\ any
visible trace of the object qualifies, rather than the fixed percentage cutoff used by
OWLv2).} Third, the skill reports predicted
tile indices and, when ground-truth labels are available, computes
accuracy. \fixme{For the evaluation reported below, predictions were instead scored
externally against a held-out answer key that the skill never had access to, avoiding the
self-graded bias of the skill's built-in accuracy check.}

\subsection{Online Challenge Solver}
Building on the offline dataset and model selection, the online solver
integrates the best-performing model and parameters for each challenge
type with additional logic to handle real-world reCAPTCHA sessions.
The online solver reuses the iframe navigation and keyword extraction
mechanism described in Section~\ref{sec:data-collection}, extended with
additional logic.

\subsubsection{Anti-Detection Measures.}
~To avoid triggering reCAPTCHA's bot-detection heuristics, the solver
implements standard anti-detection measures: hiding the
\texttt{navigator.webdriver} flag via Chrome DevTools Protocol injection,
removing Chrome automation indicators, and simulating human-like mouse
movements via \texttt{ActionChains} with small randomized offsets from the element center.

\subsubsection{Challenge Variant Detection and Solving.}
~An online reCAPTCHA session may present an unbounded sequence of
challenges before granting or denying a pass, depending on the confidence
of its risk model. The solver identifies the current challenge type and
variant by inspecting specific HTML elements: a $3{\times}3$ tile grid
indicates Type~A, while a $4{\times}4$ grid indicates Type~B. For Type~A, the sub-variant
is determined by the presence of a subtitle \texttt{<span>} element
containing ``Click verify once there are none left'': dynamic challenges
contain this element and require a loop solver that classifies tiles,
clicks matches, and waits for each tile to refresh before re-classifying;
static challenges have no such element and are solved in a single pass.
In both cases, re-clicking an already-selected tile must be avoided as
reCAPTCHA toggles selection state on repeated clicks.

After tile selection, the solver submits via
\texttt{\#recaptcha-verify-button}, which serves as the unified submit
control regardless of whether it reads ``VERIFY'', ``NEXT'', or ``SKIP.''
The solver then reads the session outcome, which falls into one of four
states: (1) \texttt{success} (checkmark granted), (2) \texttt{new\_challenge}
(a fresh challenge is presented), (3) \texttt{select\_more} (not all correct
tiles were selected), or (4) \texttt{unknown} (unexpected page state). It
continues solving until a terminal state is reached. To prevent indefinite
loops, a maximum of 20 challenge attempts per session is enforced; sessions
exceeding this limit are marked as failed.





\subsubsection{Prompt-Only Skill.}
\label{sec:prompt-only-online}
~The online prompt-only solver reuses the same tuned \texttt{/challenge-solver} skill
evaluated offline (Section~\ref{sec:prompt-only-offline}) for classification, with Selenium
handling all browser control. A self-contained script launches a Chrome browser with
anti-detection settings, navigates to a target URL, and interacts with the reCAPTCHA iframes
as described in Section~\ref{sec:data-collection}, autonomously detecting the challenge type
and sub-variant, capturing screenshots of tiles, clicking, submitting, and managing the retry loop up to
a configurable attempt limit.
The solver is
invoked via a single slash command (\texttt{/challenge-solver-online}) with a target URL,
requiring no code or configuration from the user.

\textbf{Latency Limitation.} A practical limitation of the prompt-only approach is inference
latency. Classification takes a mean of 38.4--83.7 seconds per challenge, up to 276.9 seconds
in the slowest observed case (Section~\ref{sec:eval_offline}), so sessions with multiple
sequential challenges risk expiry when classification time, tile clicking, and result checking
cumulatively approach the \emph{two-minute} limit. Expired sessions are invalidated by
reCAPTCHA, requiring the solver to restart with a fresh session and a new set of challenges.

\section{Evaluation}
\label{sec:eval}
We evaluate our solver against reCAPTCHA in two stages. First, we
report offline accuracy on our collected challenge dataset to select
the best models and parameters. Second, we evaluate end-to-end
performance against online (live) reCAPTCHA sessions.

\subsection{Offline Challenge Accuracy}
\label{sec:eval_offline}

In our offline evaluation, we define challenge accuracy as a strict all-or-nothing metric: a
challenge is counted as correct only if every correct tile is selected
and no incorrect tile is included, mirroring reCAPTCHA's own acceptance
criterion.

\textbf{CLIP on Type~A.}
Table~\ref{tab:clip} shows offline accuracy for CLIP across three model
sizes on 662 Type~A challenges.
Accuracy improves with model size but with strong diminishing returns:
a 7$\times$ increase in parameters from ViT-B/32 to ViT-H/14 yields
only a $+6$-point gain. The optimal threshold also varies across
model sizes --- ViT-H/14 requires a lower threshold (0.01) than
ViT-L/14 (0.03), reflecting that larger models produce more
discriminative embeddings with higher positive scores for true matches.
The zero-shot ceiling of approximately 58\% reflects an inherent
structural limitation: reCAPTCHA applies implicit size and visibility
thresholds when judging tile correctness, and ambiguous boundary tiles,
where only a small portion of the target
object appears, are a systematic source of error that threshold tuning
cannot resolve.

\textbf{OWLv2 vs. CLIP on Type~B.}
Type~B challenges require identifying which tiles of a single image
contain a target object. We evaluate both OWLv2 and CLIP on 338
Type~B challenges to compare object detection against per-tile
classification.
Table~\ref{tab:owlv2} shows OWLv2 accuracy at varying coverage
thresholds, with confidence fixed at 0.5.


\begin{table*}[tb]
\centering
\begin{minipage}{0.48\linewidth}
\centering
\scriptsize
\caption{CLIP zero-shot accuracy on Type~A challenges (662 challenges).}
\label{tab:clip}
\resizebox{\linewidth}{!}{%
\begin{tabular}{lrrr}
\toprule
Model & Params & Threshold & Accuracy \\
\midrule
ViT-B/32 & 86M  & 0.02 & 52\% (345/662) \\
ViT-L/14 & 304M & 0.03 & 56\% (374/662) \\
ViT-H/14 & 632M & 0.01 & \textbf{58\% (387/662)} \\
\bottomrule
\end{tabular}
}
\end{minipage}
\hfill
\begin{minipage}{0.48\linewidth}
\centering
\scriptsize
\caption{OWLv2 accuracy on Type~B challenges (338 challenges).}
\label{tab:owlv2}
\resizebox{\linewidth}{!}{%
\begin{tabular}{rrr}
\toprule
Confidence & Coverage & Accuracy \\
\midrule
0.50 &  5\% & 33\% (113/338) \\
0.50 & 10\% & \textbf{43.5\% (147/338)} \\
0.50 & 20\% & 25\% (84/338) \\
\bottomrule
\end{tabular}
}
\end{minipage}
\end{table*}

A coverage threshold of 10\% yields the best accuracy at 43.5\%. Lower
thresholds introduce false positives by selecting tiles where the
bounding box only marginally overlaps, while higher thresholds miss
tiles where the object is partially cropped at tile boundaries.

By contrast, CLIP applied per-tile achieves only 12.7\% accuracy at its
optimal threshold (0.03), a 3.4$\times$ gap compared to OWLv2.
Per-tile classification cannot determine whether a tile contains a
partial object or only background context, making it fundamentally
unsuited to Type~B's spatial reasoning demands.
Table~\ref{tab:clip_owlv2} shows the per-keyword breakdown.

\begin{table}[tb]
\caption{Per-keyword accuracy on Type~B challenges: CLIP vs.\ OWLv2
(confidence $= 0.5$, coverage $= 10\%$).}
\label{tab:clip_owlv2}
\centering
\scriptsize
\begingroup
\setlength{\tabcolsep}{4pt}
\begin{tabular}{lrrrr}
\toprule
Keyword & Count & CLIP & OWLv2 & Gap \\
\midrule
motorcycles    & 132 &  5.3\% & 35.6\% & $-$30.3pp \\
bicycles       &  68 & 20.6\% & 35.3\% & $-$14.7pp \\
traffic lights &  49 & 12.2\% & 65.3\% & $-$53.1pp \\
buses          &  33 & 27.3\% & 72.7\% & $-$45.4pp \\
crosswalks     &  24 & 12.5\% &  0.0\% & $+$12.5pp \\
stairs         &  21 &  9.5\% & 61.9\% & $-$52.4pp \\
fire hydrants  &   5 & 20.0\% & 60.0\% & $-$40.0pp \\
taxis          &   4 & 25.0\% & 75.0\% & $-$50.0pp \\
tractors       &   2 &  0.0\% & 50.0\% & $-$50.0pp \\
\midrule
\textbf{Total} & \textbf{338} & \textbf{12.7\%} & \textbf{43.5\%} & $-$30.8pp \\
\bottomrule
\end{tabular}
\endgroup
\end{table}



OWLv2 outperforms CLIP across all keywords by 14--53\%, with one exception: crosswalks, where OWLv2 fails entirely
(0.0\%) versus CLIP's 12.5\%. This is because crosswalks are flat,
ground-level patterns with no three-dimensional structure, producing
no confident detections regardless of threshold. More broadly,
performance ranges widely across keywords, from 75.0\% on taxis to
0.0\% on crosswalks. To test whether this gap could be closed through
better calibration, we additionally tuned thresholds per keyword,
yielding only marginal gains of 3--5\% and confirming that the
accuracy ceiling is structural rather than parametric.

\subsubsection{Prompt-Only Skill.}
~\fixme{On Type~A, we evaluated the \texttt{/challenge-solver} skill (Claude Sonnet~5 via
Claude Code~\cite{claude_sonnet}) on 200 challenges, independently sampled from the labeled
dataset across two separately-run batches. The skill achieved 191/200 correct (95.5\%). Every
error was within two tiles of the correct answer; seven of the nine were near-boundary
vehicle-category disagreements on ``bus'' or ``cars'' challenges (e.g., atypical vehicle shapes
such as box trucks, or a vehicle only partially in frame).}
\fixme{On Type~B, we ran the test on the set of 200 challenges. The skill achieved 101/200 correct (50.5\%).
Most Type~B errors were off by only one to three tiles at a shared boundary (a partial tile the object clips into,
or a tile whose classification depends on ambiguous conventions such as whether a second,
more distant instance of the object should count), rather than misidentifying the target
object; the sixteen-tile grid and finer spatial localization required make Type~B intrinsically
harder to get exactly right than Type~A's independent nine-tile classification, mirroring the
same gap seen with the local vision models (Table~\ref{tab:summary}).}

\textbf{Cost and Token Usage.} Each challenge is solved within a Claude Code
session, where every image read incurs the full accumulated conversation
context. \fixme{We measured this directly across the isolated invocations of both
evaluations above. For Type~A: an average of 272.0K tokens per challenge at an average cost of
\$0.150 (range \$0.079--0.553). For Type~B: an average of 616.5K tokens per challenge (roughly
2.3$\times$ Type~A's, reflecting the larger sixteen-tile grid and its higher-resolution composite
image) at an average cost of \$0.268 (range \$0.115--0.657). Both are substantially higher than
a naive single-call estimate would suggest, because even an isolated invocation is agentic:
solving one challenge requires multiple tool-use turns (reading tiles individually and, for
Type~B, reassembling a higher-resolution grid), and each turn re-incurs the system-prompt and
tool-definition overhead via the cache.}

\textbf{Latency.} \fixme{We also measured latency, taken
from each call's own reported API duration. Type~A challenges took a mean of 38.4\,s (range 20.4--166.5\,s across 200
challenges), while Type~B challenges took substantially longer at a mean of
83.7\,s (range 21.9--276.9\,s), reflecting the larger sixteen-tile grid.
These per-challenge latencies directly explain the online
session-expiration limitation discussed in Section~\ref{sec:prompt-only-online-eval}: a single
Type~B challenge alone can take over four minutes, well beyond what a live reCAPTCHA session
tolerates before timing out.}


The prompt-only skill achieves higher offline
accuracy than the corresponding local vision model on both challenge types
(\fixme{Type~A: 95.5\% vs.\ CLIP's 58\%; Type~B: 50.5\% vs.\ OWLv2's 43.5\% exact-match}), though
the margin is far narrower on Type~B, \fixme{consistent with the sixteen-tile grid's finer
spatial localization being intrinsically harder to get exactly right for either approach}. This
comes at the cost of a paid Claude subscription, \fixme{at the mean per-challenge cost reported
above}. This cost compounds with
session length: a failed session escalating to the 20-challenge limit could cost approximately \fixme{\$2--13} in
tokens. The higher per-challenge accuracy partially offsets this, however, by reducing the
average number of challenges needed per successful session.


\subsection{Online Session Accuracy}

\subsubsection{Local Vision Models.}
\begin{figure}[tbh]
\centering
\includegraphics[width=\columnwidth]{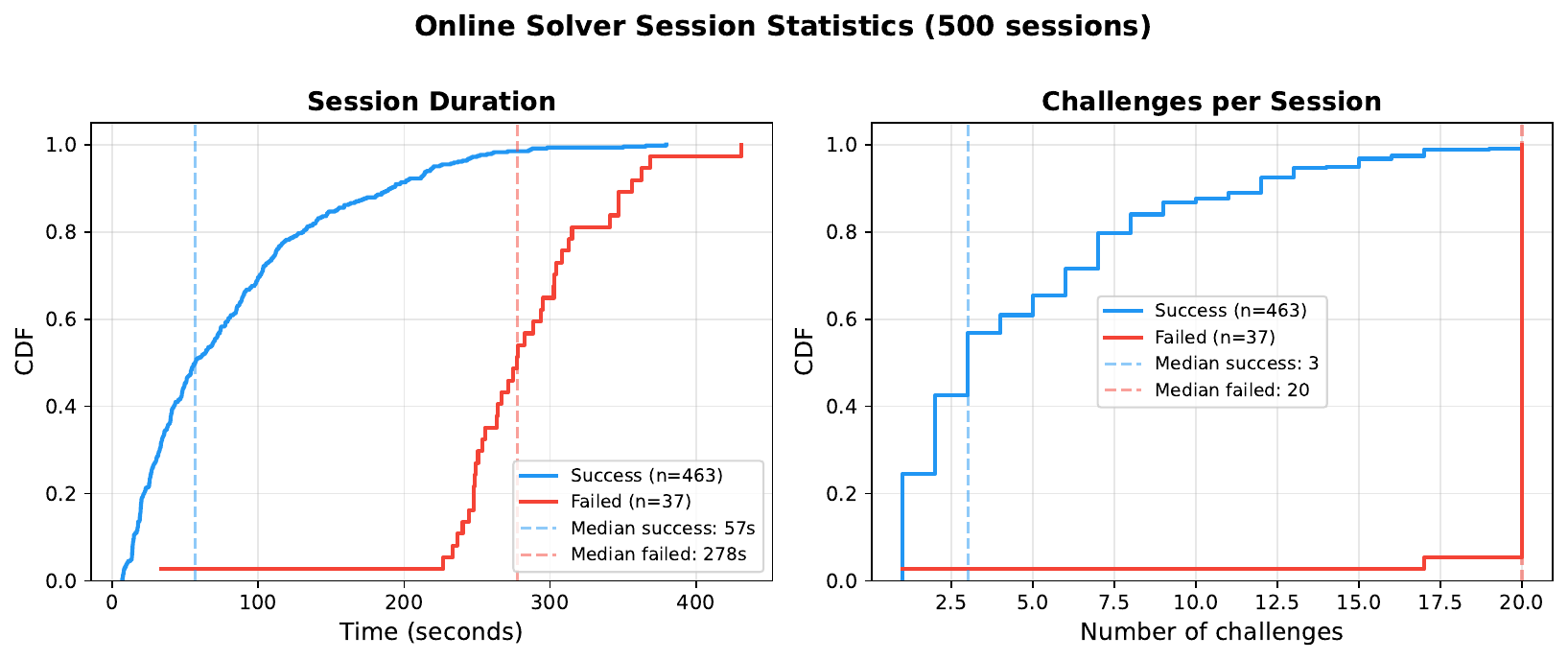}
\caption{CDF of session duration (left) and number of challenges per
session (right), split by outcome. Successful sessions complete with a
median of 57s and 3 challenges; failed sessions cluster at the
20-challenge limit (median 278s).}
\label{fig:online_cdf}
\end{figure}

~We ran the fully automated solver across 500 real-world reCAPTCHA
sessions including the official demo page, third-party demo
sites, and live production login pages, with a maximum of 20 challenge attempts per session.
Table~\ref{tab:online} summarizes the results.
Across 2,931 challenges, 67\% were Type~B and 33\% were Type~A.
Figure~\ref{fig:online_cdf} shows the cumulative distribution of session
duration and challenge count, split by outcome.

\begin{table}[tb]
\caption{Online solver results (500 sessions, 2,931 challenges).}
\label{tab:online}
\centering
\scriptsize
\begingroup
\setlength{\tabcolsep}{4pt}
\begin{tabular}{lrr}
\toprule
 & Successful (463) & Failed (37) \\
\midrule
Session success rate    & \textbf{92.6\% (463/500)}  & \textbf{7.4\%} \\
Avg.\ challenges per session & 4.8 & 19.4 \\
Min / Median / Max challenges per session & 1 / 3 / 20 & 20 / 20 / 20 \\
Avg.\ time & 81s & 282s \\
Min / Median / Max time & 7s / 57s / 380s & 33s / 278s / 431s \\
\bottomrule
\end{tabular}
\endgroup
\end{table}

The solver achieves a 92.6\% session success rate across 500 sessions.
Successful sessions complete quickly, with a median of 3 challenges and
57 seconds. Failed sessions, by contrast, have a median of 20 challenges
and 278 seconds, indicating that failures are not caused by
misclassification but by reCAPTCHA's risk model sustaining suspicion
and serving challenges until the attempt limit is exhausted.

\textbf{Type~B as the Primary Failure Driver.}
To understand what distinguishes failed sessions from successful ones,
we analyze the mix of challenge types presented per session.
Table~\ref{tab:typemix} shows session success rate stratified by the
proportion of Type~B challenges received.
The results confirm that Type~B is the primary failure driver. Sessions
consisting entirely of Type~A challenges succeed at 99.5\%, effectively
perfect. As the proportion of Type~B challenges increases, success rate
drops markedly: sessions with 50\% or more Type~B challenges succeed at
only 85.8--88.4\%. This is consistent with OWLv2's offline accuracy of
43.5\% on Type~B, a difference that
directly links offline model accuracy to real-world session outcomes.

\begin{table}[tb]
\caption{Session success rate by Type~B challenge proportion.}
\label{tab:typemix}
\centering
\scriptsize
\begingroup
\setlength{\tabcolsep}{4pt}
\begin{tabular}{lrr}
\toprule
Type~B proportion & Sessions & Success rate \\
\midrule
0\% (all Type~A)   & 218 & 99.5\% \\
1--49\% Type~B     &  21 & 100.0\% \\
50--99\% Type~B    & 218 & 85.8\% \\
100\% (all Type~B) &  43 & 88.4\% \\
\bottomrule
\end{tabular}
\endgroup
\end{table}

\subsubsection{Prompt-Only Skill.}
\label{sec:prompt-only-online-eval}
~The online prompt-only solver drives the browser via Selenium and reuses the same tuned
\texttt{/challenge-solver} skill evaluated above for classification (Section~\ref{sec:prompt-only-online}).
In verification against the live reCAPTCHA demo page, sessions consisting of a single
static Type~A or Type~B challenge succeeded consistently. However, sessions with dynamic
Type~A challenges or high-volume Type~B challenges expose a real latency limitation:
\fixme{inference and tile classification take 38.4--83.7 seconds on average per challenge, and
up to 276.9 seconds in the slowest observed case (Section~\ref{sec:eval_offline})}; and dynamic
challenges require multiple tile-refresh cycles, causing sessions to expire before completion.
Notably, every failure observed in this verification was a session timeout, not a
misclassification: the skill's predicted tiles were correct in every case observed, but the
cumulative classification time exceeded reCAPTCHA's session window before the solver could
submit. Given this latency sensitivity and the
small scale of verification conducted, we do not report a formal online session success
rate for the prompt-only solver. The local vision model solver, by contrast, is unaffected by
this limitation, classifying tiles in under one second.

\subsection{\fixme{Computational Cost}}
\label{sec:compute-cost}
\fixme{To make our ``zero-cost'' claim precise, Table~\ref{tab:compute-cost} reports the
storage, peak memory, and per-challenge latency of the two deployed models. Both fit in a
few gigabytes and solve a challenge in under one second on an integrated GPU; crucially,
both also run CPU-only, so no GPU is required. The solver thus incurs no API or per-query
charges: its only costs are a few gigabytes of storage and, at an estimated 10--15\,J per
challenge, negligible electricity.}
\begin{table}[tb]
\centering
\scriptsize
\caption{\fixme{Computational cost of the two deployed local models, measured on a
consumer laptop.}}
\label{tab:compute-cost}
\begingroup
\setlength{\tabcolsep}{4pt}
\begin{tabular}{lrrrr}
\toprule
Model (challenge type) & Storage & Peak RAM & Latency (GPU) & Latency (CPU) \\
\midrule
CLIP ViT-H/14 (Type~A) & 3.7\,GB & 4.3\,GB & \fixme{0.70\,s} & 2.6\,s \\
OWLv2 base (Type~B)    & 591\,MB & 1.2\,GB & \fixme{0.52\,s} & 1.0\,s \\
\bottomrule
\end{tabular}
\endgroup
\end{table}

\subsection{Summary}
Table~\ref{tab:summary} summarizes all evaluated approaches across both
offline and online settings. Note that offline and online accuracy
measure different things: offline accuracy is per-challenge, a single
CAPTCHA with one correct answer set, while online accuracy is
per-session, where each session may contain multiple challenges that
must all be resolved before reCAPTCHA grants a pass. The offline
setting thus isolates pure model capability, while the online setting
better mirrors a real-world attacker scenario.





\begin{table}[tb]
\caption{Solver comparison across offline (per-challenge) and online (per-session).}
\label{tab:summary}
{
\scriptsize
\centering
\resizebox{\linewidth}{!}{%
\begingroup
\setlength{\tabcolsep}{4pt}
\begin{tabular}{p{2.6cm}p{0.6cm}lllp{2.6cm}}
\toprule
Method & Type & Accuracy & Avg. Cost & Avg. Latency & Approach \\
\midrule
\multicolumn{6}{l}{\textit{Offline (per-challenge accuracy)}} \\
\midrule
CLIP ViT-H/14  & A   & 58.0\%            & \$0           & \fixme{0.70\,s} & Per-tile classification \\
OWLv2          & B   & 43.5\%            & \$0           & \fixme{0.52\,s} & Object detection \\
\fixme{Prompt-only skill} & \fixme{A} & \fixme{95.5\%$^*$} & \fixme{\$0.079--0.553$^\dagger$} & \fixme{38.4\,s} & \fixme{Multimodal vision} \\
\fixme{Prompt-only skill} & \fixme{B} & \fixme{50.5\%$^\S$} & \fixme{\$0.115--0.657$^\dagger$} & \fixme{83.7\,s} & \fixme{Multimodal vision} \\
\midrule
\multicolumn{6}{l}{\textit{Online (per-session accuracy)}} \\
\midrule
CLIP+OWLv2 & A+B & 92.6\%             & \$0            & \fixme{81\,s$^\P$} & Per-tile classification \\
    & & & & & or Object detection\\
Prompt-only skill & A+B & N/A$^\ddagger$ & N/A$^\ddagger$ & \fixme{N/A$^\ddagger$} & Multimodal vision \\
\bottomrule
\end{tabular}
\endgroup
}
}

{\scriptsize
$^*$\fixme{Sampled 200 Type~A challenges; see Section~\ref{sec:prompt-only-offline}}.\\
$^\S$\fixme{Sampled 200 Type~B challenges.}\\
$^\ddagger$Online results are not reported due to latency sensitivity; see Section~\ref{sec:prompt-only-online-eval}.\\
$^\dagger$Requires Claude AI subscription.\\
$^\P$\fixme{Mean time for successful sessions; failed sessions have a mean of 282\,s (Table~\ref{tab:online}).
}
}
\end{table}



\textbf{Interpreting Offline and Online Accuracy.}
The online session success rate (92.6\%) appears higher than offline
per-challenge accuracy (58\% Type~A, 43.5\% Type~B), which warrants
explanation. Offline accuracy uses strict exact-match evaluation: a
challenge is correct only if every correct tile is selected and no
incorrect tile is included. In practice, however, reCAPTCHA applies
more relaxed acceptance criteria that account for human ambiguity, and
this leniency benefits attackers as much as legitimate users. For
example, a challenge instructing users to ``Select all images with
motorcycles'' may accept selections that include the human rider, as
these reflect natural human interpretation. As a result, our offline
accuracy figures represent conservative lower bounds on real-world
performance: the solver succeeds online in cases our strict offline
metric would count as failures.

\section{Implications for CAPTCHA Security}
\label{sec:discussion}

\subsection{Bots Now Always Win.}
Visual CAPTCHAs operate under the assumption that machines cannot
reliably identify objects in images at human levels. Our results, along
with recent work, show this assumption is no longer valid. \fixme{As an
illustrative estimate of the retry advantage, assume attempts are independent
and that a single challenge success within the attempt budget is sufficient:
even our weakest model, at 43.5\% per-challenge accuracy, would then succeed
with probability $1 - (1-0.435)^5 \approx 94.3\%$ within five attempts, rising
to $1 - (1-0.58)^5 \approx 98.7\%$ at 58.0\% accuracy. We treat these values as
a simplified model rather than a measurement of session success; the realized
per-session success rate under reCAPTCHA's actual retry and risk-scoring
behavior is measured directly in our online evaluation (92.6\%,
Section~\ref{sec:eval}).} Combined with Sivakorn
et al.~\cite{sivakorn2016} (70.78\% in 2016) and
Halligan~\cite{halligan2025} (60.7\% across 26 types), the evidence
is consistent: visual image challenges are no longer bot-hard.

\textbf{Zero-Code Barrier.} Perhaps the most alarming finding is that
reCAPTCHA can be broken without any technical expertise. Our prompt-only
approach requires no code, no machine learning knowledge, and no external
services, just natural-language instructions.
This reduces the barrier to entry to near zero: any user with access to
an AI assistant can construct a working CAPTCHA solver.

\textbf{Future of CAPTCHA Security.}
As AI capabilities become increasingly accessible, the gap between what
machines and humans can do on challenge-based tasks is rapidly closing.
The industry is already moving in the right direction: reCAPTCHA v3 and
similar behavior-based systems avoid visual challenges entirely for most
users. However, our results expose a critical weakness in this
transition: the challenge-based fallback for low-reputation sessions
remains active and, as we demonstrate, fully exploitable. Ironically,
this fallback may attract attackers more than v3 alone, since a bot that
fails behavioral analysis is escalated directly to a visual challenge,
creating a predictable and reliable attack path.

\subsection{Responsible Disclosure and Ethics}
\fixme{We disclosed these findings to Google prior to camera-ready publication, through
Google's Bug Hunters, on September 6, 2026 (issue ID:
557908167).
All experiments were conducted on fresh, unauthenticated browser sessions with no Google sign-in
and no user data, so no user account was targeted or affected, and we treat the study as a
defensive measurement of a deployed control under worst-case, low-reputation conditions.
To limit misuse, we do not release the end-to-end online solver, which automates browser interaction to carry out the full attack against a live challenge.}


\section{Reproducibility Statement}


The labeled challenge dataset is publicly available to support 
reproducibility and further benchmarking at 
\fixme{\url{https://github.com/ssivakorn/reCAPTCHA-study}, and mirrored on Kaggle at
\url{https://www.kaggle.com/datasets/ssivakorn/recaptcha-challenge}}.
\fixme{This repository also contains the natural-language classification prompt (the
\texttt{challenge-solver} skill) used for our prompt-only evaluation. The prompt-only runs used Claude
Sonnet~5 via Claude Code~\cite{claude_sonnet} in its default configuration, with extended
thinking not enabled.
}
Video demonstrations of the online solver on live reCAPTCHA sessions 
are available at 
\url{https://youtube.com/playlist?list=PLqZ9f4-RC3S-2UaC4WW456xvKInqHeN8_&si=Yqdl-Til6bVPJapn}.

\section{Conclusion}
\label{sec:conclusion}




We presented a comprehensive study of Google reCAPTCHA's visual image
challenges, introducing a taxonomy of two challenge types: Type~A
(independent image tiles) and Type~B (single image partitioned into a
grid), each requiring a fundamentally different solving strategy. Using
only free, open-source, locally-run models, we achieved 58.0\%
per-challenge accuracy on Type~A with CLIP and 43.5\% on Type~B with
OWLv2, with no training and no API cost. An end-to-end automated solver
built on these models achieves a 92.6\% session success rate across 500
real-world reCAPTCHA sessions.

Beyond the specific case of reCAPTCHA, our findings carry broader
implications. The bot-hard assumption that visual recognition tasks or
any task-based challenge reliably separates humans from machines no
longer holds. Compounding this, we show that effective solvers require
neither code nor technical expertise: a non-technical adversary armed
only with a commodity AI assistant and a natural-language instruction
can defeat the system entirely. Together, these results suggest that
challenge-based verification has reached an inflection point from which
it is unlikely to recover. We hope this work motivates the security
community to accelerate the transition toward verification mechanisms
that do not rely on task difficulty as a security primitive.

\begin{credits}
\subsubsection{\ackname}
\fixme{We would like to thank the anonymous reviewers, as well as Chaim Haas and
the Intellectual Property Working Group (IPWG) team at Bloomberg, for their careful
review and valuable feedback that improved this manuscript's clarity.}

\subsubsection{\discintname}
\fixme{The authors have no competing interests to declare that are relevant to the content of
this article.}
\end{credits}

\bibliographystyle{splncs04}
\bibliography{references}

@inproceedings{sivakorn2016,
  author    = {Sivakorn, Suphannee and Polakis, Iasonas and Keromytis, Angelos D.},
  title     = {{I Am Robot: (Deep) Learning to Break Semantic Image CAPTCHAs}},
  booktitle = {Proceedings of the 2016 IEEE European Symposium on Security and Privacy (EuroS\&P)},
  year      = {2016},
  pages     = {388--403},
  publisher = {IEEE},
  doi       = {10.1109/EuroSP.2016.37}
}

@inproceedings{halligan2025,
  author    = {Teoh, Xiwen and Lin, Yun and Li, Siqi and Liu, Ruofan and Sollomoni, Avi and Harel, Yaniv and Dong, Jin Song},
  title     = {{Are CAPTCHAs Still Bot-hard? Generalized Visual CAPTCHA Solving with Agentic Vision Language Model}},
  booktitle = {Proceedings of the 34th USENIX Security Symposium},
  year      = {2025},
  pages     = {3747--3766},
  publisher = {USENIX Association}
}

@misc{radford2021clip,
  title={Learning Transferable Visual Models From Natural Language Supervision}, 
  author={Alec Radford and Jong Wook Kim and Chris Hallacy and Aditya Ramesh and Gabriel Goh and Sandhini Agarwal and Girish Sastry and Amanda Askell and Pamela Mishkin and Jack Clark and Gretchen Krueger and Ilya Sutskever},
  year={2021},
  eprint={2103.00020},
  archivePrefix={arXiv},
  primaryClass={cs.CV},
  url={https://arxiv.org/abs/2103.00020}
}

@inproceedings{minderer2023owlv2,
  author    = {Minderer, Matthias and Gritsenko, Alexey and Houlsby, Neil},
  title     = {{Scaling Open-Vocabulary Object Detection}},
  booktitle = {Advances in Neural Information Processing Systems (NeurIPS)},
  year      = {2023},
  pages     = {72983--73007},
  volume    = {36}
}

@INPROCEEDINGS{kirillov2023sam,
  author={Kirillov, Alexander and Mintun, Eric and Ravi, Nikhila and Mao, Hanzi and Rolland, Chloe and Gustafson, Laura and Xiao, Tete and Whitehead, Spencer and Berg, Alexander C. and Lo, Wan-Yen and Dollár, Piotr and Girshick, Ross},
  booktitle={2023 IEEE/CVF International Conference on Computer Vision (ICCV)}, 
  title={Segment Anything}, 
  year={2023},
  volume={},
  number={},
  pages={3992-4003},
  doi={10.1109/ICCV51070.2023.00371}}

@ARTICLE{10.3389/fnins.2017.00309,
    AUTHOR={Li, Hongmin  and Liu, Hanchao  and Ji, Xiangyang  and Li, Guoqi  and Shi, Luping },
    TITLE={{CIFAR10-DVS: An Event-Stream Dataset for Object Classification}},
    JOURNAL={Frontiers in Neuroscience},
    VOLUME={11},
    YEAR={2017},
    URL={https://www.frontiersin.org/journals/neuroscience/articles/10.3389/fnins.2017.00309},
    DOI={10.3389/fnins.2017.00309},
    ISSN={1662-453X}
}

@article{ILSVRC15,
Author = {Olga Russakovsky and Jia Deng and Hao Su and Jonathan Krause and Sanjeev Satheesh and Sean Ma and Zhiheng Huang and Andrej Karpathy and Aditya Khosla and Michael Bernstein and Alexander C. Berg and Li Fei-Fei},
Title = {{ImageNet Large Scale Visual Recognition Challenge}},
Year = {2015},
journal   = {International Journal of Computer Vision (IJCV)},
doi = {10.1007/s11263-015-0816-y},
volume={115},
number={3},
pages={211-252}
}

@INPROCEEDINGS{8887344,
  author={Zhang, Yang and Gao, Haichang and Pei, Ge and Luo, Sainan and Chang, Guoqin and Cheng, Nuo},
  booktitle={2019 18th IEEE International Conference On Trust, Security And Privacy In Computing And Communications/13th IEEE International Conference On Big Data Science And Engineering (TrustCom/BigDataSE)}, 
  title={A Survey of Research on CAPTCHA Designing and Breaking Techniques}, 
  year={2019},
  volume={},
  number={},
  pages={75-84},
  doi={10.1109/TrustCom/BigDataSE.2019.00020}}

@ARTICLE{8665729,
  author={Weng, Haiqin and Zhao, Binbin and Ji, Shouling and Chen, Jianhai and Wang, Ting and He, Qinming and Beyah, Raheem},
  journal={Big Data Mining and Analytics}, 
  title={Towards understanding the security of modern image captchas and underground captcha-solving services}, 
  year={2019},
  volume={2},
  number={2},
  pages={118-144},
  doi={10.26599/BDMA.2019.9020001}}

@inproceedings{dosovitskiy2020vit,
    title={An Image is Worth 16x16 Words: Transformers for Image Recognition at Scale},
    author={Alexey Dosovitskiy and Lucas Beyer and Alexander Kolesnikov and Dirk Weissenborn and Xiaohua Zhai and Thomas Unterthiner and Mostafa Dehghani and Matthias Minderer and Georg Heigold and Sylvain Gelly and Jakob Uszkoreit and Neil Houlsby},
    booktitle={International Conference on Learning Representations},
    year={2021},
    url={https://openreview.net/forum?id=YicbFdNTTy}
}

@inproceedings{redmon2016yolo,
  author    = {Redmon, Joseph and Divvala, Santosh and Girshick, Ross and Farhadi, Ali},
  title     = {{You Only Look Once: Unified, Real-Time Object Detection}},
  booktitle = {Proceedings of the IEEE Conference on Computer Vision and Pattern Recognition (CVPR)},
  year      = {2016},
  pages     = {779--788}
}

@misc{friendlycaptcha2026,
  author       = {Mitifiot, Pauline},
  organization = {{Friendly Captcha}},
  title        = {{reCAPTCHA v2 vs. v3: Effective Bot Protection?}},
  year         = {2026},
  date         = {7},
  url          = {https://friendlycaptcha.com/insights/recaptcha-v2-vs-v3/}
}

@misc{glasswing2025,
  author       = {{Anthropic}},
  organization = {{Anthropic}},
  title        = {{Expanding Project Glasswing}},
  year         = {2026},
  howpublished = {\url{https://www.anthropic.com/news/expanding-project-glasswing}}
}

@misc{ft2025ai,
  author       = {Criddle, Cristina and Sevastopulo, Demetri},
  organzation  = {{Financial Times}},
  title        = {{US National Security Agency using Anthropic's Mythos for cyber attacks}},
  year         = {2026},
  howpublished = {\url{https://www.ft.com/content/d02d91b3-2636-454e-9442-dc7e69f51815}}
}

@misc{icann2023,
  author       = {Tajalizadehkhoob, Samaneh},
  organization = {{ICANN}},
  title        = {New {ICANN} Project Explores the Drivers of Malicious Domain Name Registrations},
  year         = {2023},
  howpublished = {\url{https://www.icann.org/en/blogs/details/new-icann-project-explores-the-drivers-of-malicious-domain-name-registrations-25-04-2023-en}}
}

@misc{falconfeeds2024,
  author       = {{FalconFeeds}},
  organization = {{FalconFeeds}},
  title        = {{The Life Cycle of a Malicious Domain: From Registration to Takedown}},
  year         = {2025},
  howpublished = {\url{https://falconfeeds.io/blogs/malicious-domain-life-cycle-analysis/}}
}

@misc{cloudflare2022,
  author       = {Wolters, Benedikt and Guerreiro, Maxime and Martinetti, Adam},
  organization = {{Cloudflare}},
  title        = {Turnstile {GA}: A user-friendly {CAPTCHA} alternative},
  year         = {2023},
  howpublished = {\url{https://blog.cloudflare.com/turnstile-ga/}}
}

@misc{recaptchav2,
    author = {{Google}},
    organization = {{Google}},
    title = {{reCAPTCHA v2}},
    howpublished = {\url{https://developers.google.com/recaptcha/docs/display}}
}

@misc{recaptchav3,
  author       = {{Google}},
  organization = {{Google}},
  title        = {{reCAPTCHA} v3},
  howpublished = {\url{https://developers.google.com/recaptcha/docs/v3}}
}

@misc{thales2026badbots,
  author       = {{imperva}},
  organization = {{imperva}},
  title        = {{2026 Bad Bot Report: Bad Bots in the Agentic Age}},
  year         = {2026},
  howpublished = {\url{https://www.imperva.com/resources/resource-library/reports/2026-bad-bot-report/}}
}

@inproceedings{bursztein2011text,
  author    = {Bursztein, Elie and Martin, Matthieu and Mitchell, John},
  title     = {Text-based {CAPTCHA} Strengths and Weaknesses},
  booktitle = {Proceedings of the 18th ACM Conference on Computer and 
               Communications Security (CCS)},
  year      = {2011}
}

@misc{chen2021codex,
  author={Mark Chen and Jerry Tworek and Heewoo Jun and Qiming Yuan and Henrique Ponde de Oliveira Pinto and Jared Kaplan and Harri Edwards and Yuri Burda and Nicholas Joseph and Greg Brockman and Alex Ray and Raul Puri and Gretchen Krueger and Michael Petrov and Heidy Khlaaf and Girish Sastry and Pamela Mishkin and Brooke Chan and Scott Gray and Nick Ryder and Mikhail Pavlov and Alethea Power and Lukasz Kaiser and Mohammad Bavarian and Clemens Winter and Philippe Tillet and Felipe Petroski Such and Dave Cummings and Matthias Plappert and Fotios Chantzis and Elizabeth Barnes and Ariel Herbert-Voss and William Hebgen Guss and Alex Nichol and Alex Paino and Nikolas Tezak and Jie Tang and Igor Babuschkin and Suchir Balaji and Shantanu Jain and William Saunders and Christopher Hesse and Andrew N. Carr and Jan Leike and Josh Achiam and Vedant Misra and Evan Morikawa and Alec Radford and Matthew Knight and Miles Brundage and Mira Murati and Katie Mayer and Peter Welinder and Bob McGrew and Dario Amodei and Sam McCandlish and Ilya Sutskever and Wojciech Zaremba},
  title   = {{Evaluating Large Language Models Trained on Code}},
  year = {2021},
  eprint={2107.03374},
  archivePrefix={arXiv},
  primaryClass={cs.LG},
  url={https://arxiv.org/abs/2107.03374}, 
}

@misc{claude_multimodal,
    author       = {{Stream.io, Inc.}},
    organization = {{Stream.io, Inc.}},
    title        = {{Visual Intelligence in Claude: Interpreting Documents and Structured Content}},    
    year         = {2026},
    howpublished = {\url{https://getstream.io/blog/anthropic-claude-visual-reasoning/}} 
}

@misc{claude_sonnet,
    author       = {{Anthropic}},
    organization = {{Anthropic}},
    title = {{Introducing Claude Sonnet 5}},
    year = 2026,
    howpublished = {\url{https://www.anthropic.com/news/claude-sonnet-5}}
}

@misc{claude_code,
    author       = {{Anthropic}},
    organization = {{Anthropic}},
    title = {{Claude Code by Anthropic | AI Coding Agent, Terminal, IDE}},
    howpublished = {\url{https://claude.ai/code}},

}

@misc{openai2024gpt4o,
    author       = {{OpenAI}},
    organization = {{OpenAI}},
    title        = {{Hello GPT-4o}},
    year         = {2024},
    howpublished = {\url{https://openai.com/index/hello-gpt-4o/}}
}

@misc{openai2025gpt5,
    author = {{OpenAI}},
    organization = {{OpenAI}},
    title  = {{Introducing GPT‑5}},
    year   = {2025},
    howpublished = {\url{https://openai.com/index/introducing-gpt-5/}}

}

@misc{google2024gemini,
    author       = {{Google DeepMind}},
    organization = {{Google}},
    title        = {{Gemini: Frontier intelligence with action}},
    year         = {2026},
    howpublished = {\url{https://deepmind.google/models/gemini/}}
}

@inproceedings{liu2023llava,
    author      = {Liu, Haotian and Li, Chunyuan and Wu, Qingyang and Lee, Yong Jae},
    title       = {Visual Instruction Tuning},
    booktitle   = {NeurIPS},
    year        = {2023}
}

@misc{chen2024internvl,
      title={InternVL: Scaling up Vision Foundation Models and Aligning for Generic Visual-Linguistic Tasks}, 
      author={Zhe Chen and Jiannan Wu and Wenhai Wang and Weijie Su and Guo Chen and Sen Xing and Muyan Zhong and Qinglong Zhang and Xizhou Zhu and Lewei Lu and Bin Li and Ping Luo and Tong Lu and Yu Qiao and Jifeng Dai},
      year={2024},
      eprint={2312.14238},
      archivePrefix={arXiv},
      primaryClass={cs.CV},
      url={https://arxiv.org/abs/2312.14238}, 
}

@inproceedings{deng2024pentestgpt,
  title     = {PentestGPT: Evaluating and Harnessing Large Language
               Models for Automated Penetration Testing},
  author    = {Deng, Gelei and Liu, Yi and Mayoral-Vilches, Víctor and
               Liu, Peng and Li, Yuekang and Xu, Yuan and Zhang, Tianwei
               and Liu, Yang and Pinzger, Martin and Rass, Stefan},
  booktitle = {33rd USENIX Security Symposium},
  year      = {2024},
  address   = {Philadelphia, PA},
  publisher = {USENIX Association}
}

@misc{fang2024llmagents,
      title={LLM Agents can Autonomously Exploit One-day Vulnerabilities}, 
      author={Richard Fang and Rohan Bindu and Akul Gupta and Daniel Kang},
      year={2024},
      eprint={2404.08144},
      archivePrefix={arXiv},
      primaryClass={cs.CR},
      url={https://arxiv.org/abs/2404.08144}, 
}

@misc{selenium,
    title = {{Selenium Automates Browsers}},
    organization = {{Software Freedom Conservancy}},
    author = {{Selenium}},
    howpublished={\url{https://www.selenium.dev/}}
}

@misc{chromedriver,
    title = {{What is ChromeDriver?}},
    author = {{Chrome for Developers}},
    organization = {{Google}},
    year = {2025},
    howpublished = {\url{https://developer.chrome.com/docs/chromedriver/}},

}

@INPROCEEDINGS {10633630,
    author = { Plesner, Andreas and Vontobel, Tobias and Wattenhofer, Roger },
    booktitle = { 2024 IEEE 48th Annual Computers, Software, and Applications Conference (COMPSAC) },
    title = {{ Breaking reCAPTCHAv2 }},
    year = {2024},
    volume = {},
    ISSN = {},
    pages = {1047-1056},
    doi = {10.1109/COMPSAC61105.2024.00142},
    url = {https://doi.ieeecomputersociety.org/10.1109/COMPSAC61105.2024.00142},
    publisher = {IEEE Computer Society},
    address = {Los Alamitos, CA, USA},
    month =Jul
}





\end{document}